\documentclass[twocolumn]{aa}
\usepackage{graphicx}
\usepackage{txfonts}
\begin{document}
   \title{High-Energy sources before INTEGRAL}

   \subtitle{INTEGRAL reference catalog}

   \author{Ken Ebisawa\inst{1}$^,$\inst{2}\fnmsep\thanks{Universities Space Research Association}
          \and
          G. Bourban\inst{3}
	  \and
          A. Bodaghee\inst{1,3}
	  \and
          N. Mowlavi\inst{1,3}
	  \and
          T. J.-L. Courvoisier\inst{1,3}
          }

   \offprints{Ken Ebisawa}

   \institute{INTEGRAL Science Data Center, 16 chemin d'Ecogia, 1290 Versoix, Switzerland\\
              \email{ebisawa@obs.unige.ch}
         \and
             code 662, NASA/GSFC, Greenbelt, 20771 Maryland, USA
         \and
             Observatory of Geneva, 51 chemin des Maillettes, 1290 Sauverny, Switzerland\\
             }

   \date{Received June 99, 2003; accepted July 99, 2003}

   \abstract{
     We describe the INTEGRAL reference catalog which classifies previously known 
     bright X-ray and gamma-ray sources before the launch of INTEGRAL.
     These sources are, or have been at least once, brighter than $\sim1 $ mCrab above 3 keV,
     and are expected to be detected by INTEGRAL.
     This catalog is being used in the INTEGRAL {\it Quick Look Analysis}\/ to discover new sources or
     significantly variable sources.
     We compiled several published X-ray and gamma-ray catalogs, and surveyed recent
     publications for new sources.
     Consequently, there   are 1122 sources in our INTEGRAL reference catalog.  In addition
     to the source positions, we show an approximate spectral model and expected flux for each source, based on 
     which we derive expected INTEGRAL counting rates.  Assuming the
     default instrument performances and 
     at least $\sim 10^5$ sec exposure
     time for any part of the sky, we expect that INTEGRAL will detect
     at least $\sim$ 700 sources below 10 keV and $\sim$ 400 sources above 20 keV  over the mission life.
   }

   \maketitle
%

   \begin{figure*}
   \centering
   \includegraphics[angle=-90,width=16cm]{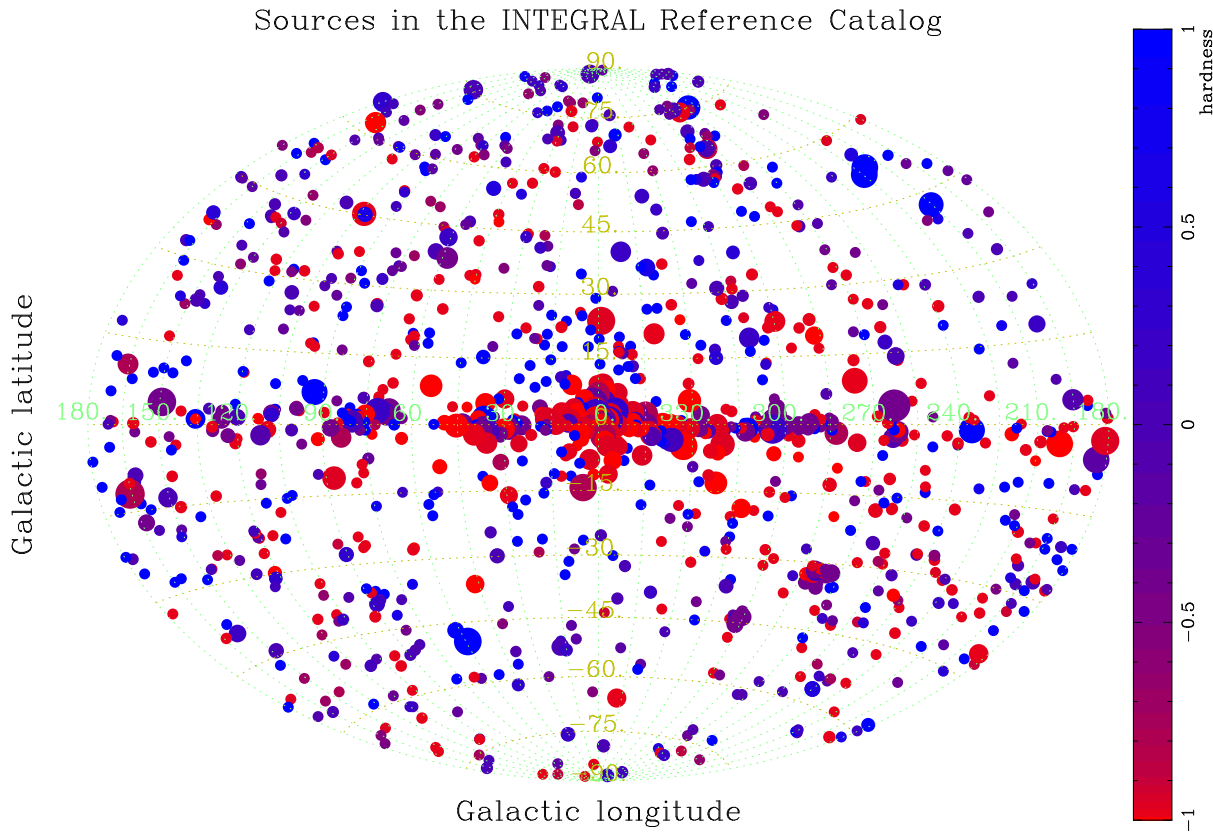}
      \caption{All the sources included in the INTEGRAL reference catalog.
              Size and color of the symbols indicate approximate source brightness and spectral
              hardness, respectively.  Symbol size is logarithmically proportional to the
              total JEMX + ISGRI counting rates. The hardenss is defined as $H-S/H+S$, where
              $H$ is the total ISGRI counting rates (20 -- 200 keV) 
      and $S$ is the soft-band (3 -- 10 keV)
              JEMX counting rates.
              }
         \label{Figure}
   \end{figure*}

\section{Introduction}

One of the main purposes of the INTEGRAL mission is to discover new
X-ray/gamma-ray sources, and also to monitor significantly variable sources.
When new or transient sources are detected, INTEGRAL is
expected to give prompt announcements  to the worldwide astronomical community.
Near real-time monitoring of the INTEGRAL data, i.e.,  {\em Quick Look Analysis}
(QLA), is carried out at  the INTEGRAL Science Data Center (ISDC;
Courvoisier et al.\ 2003a).
To this end, we need a reliable X-ray/gamma-ray source catalog with which
INTEGRAL QLA results can be compared.

Many X-ray/gamma-ray source catalogs have already been published.  
To make the  {\em INTEGRAL Reference Catalog},
which we describe in the current paper, 
we searched the literature,  took several high-energy catalogs,
combined them and selected sources to suit our needs. 
Note that the INTEGRAL Reference Catalog is continuously
updated for the QLA purpose by incorporating new sources
discovered by INTEGRAL and other high-energy satellites.
In this paper, we describe all the sources
included in the INTEGRAL Reference Catalog
 at the end of June 2003, {\em except}\/
those sources discovered by INTEGRAL.

All the sources in the INTEGRAL Reference Catalog are shown in Figure 1
in Galactic coordinates. Size and color of the symbol indicate
approximate fluxes  and spectral hardness.

\section{Method of selecting sources}

The idea behind the  INTEGRAL Reference Catalog is to put all sources which
are {\em possibly}\/ detected and identified by INTEGRAL.  Corresponding to 
the expected sensitivity and energy range of INTEGRAL,
we select sources  that are, or have been at least once,  
brighter than $\sim$ 1 mCrab 
in any energy band from 3 keV to 8000 MeV (energy range
covered with JEMX, IBIS and SPI). Thus, for example, 
historic X-ray novae are included even if they are expected
in quiescence when observed with INTEGRAL, since these sources {\em may}\/ be
detected when brightened.  In addition, we include
all the EGRET sources in Macomb and Gehrels (1999) even if they have never been
detected in the INTEGRAL energy band.

We took all the sources in the following catalogs.
\begin{itemize}
\item  High-mass X-ray binary catalog by Liu et al.\ (2000).
\item  Low-mass X-ray binary catalog by Liu et al.\ (2001).
\item   Gamma-ray sources catalog by Macomb and Gehrels (1999).
\item The 4th  Uhuru catalog by Forman et al.\ (1978).
\item  HEAO1 A4 catalog by Levine et al.\ (1984).
\item  BATSE observation of the Piccinotti's sample AGNs.
\item  ASCA AGN data from the Tartarus database (http://tartarus.gsfc.nasa.gov).
\item  IAUC circulars until June 30, 2003.
\end{itemize}

\section{Information included in the catalog}

The INTEGRAL Reference Catalog is published in the electronic form
together with the present paper.  It is an ASCII file written in the \LaTeX\/ tabular form,
so that readers can run \LaTeX\/ and  easily print.  Since column entries are formatted
with fixed-width columns, the file is legible even in the original ASCII format.

The catalog is also available in the html version and FITS version from the
ISDC web-site ({\tt http://isdc.unige.ch}).  The FITS version of the catalog is
used in the actual QLA operation and INTEGRAL data analysis.  In the html version,
the source name and the position reference is hyper-linked to its corresponding page in SIMBAD
and ADS, respectively.

The following is an explanation of each column in the catalog:

\begin{itemize}
\item {\em Catalog source number}  
\item {\em Source name}:  \\Often there are two or more names commonly used for the
      same source.  We take the name which we consider  most
      popular in high-energy astrophysics and is also used
      as a SIMBAD source identifier.  Thus, users can directly put
      the source name into the SIMBAD query box.
\item {\em Right Ascension and Declination in J2000 coordinates}: \\
     These are given in ``hh mm ss.s'' and ``deg arcmin arcsec''
     format, respectively.
     In most cases, positions are taken from SIMBAD.
     From time to time, however, 
     we found and adopted a more recent position determination than SIMBAD.
     Accuracy depends on individual sources, which is reflected in the
     representation of the source positions.  For example, 
     an Uhuru source 4U 1130$-$14 has  the source position (11 33, $-$14.9), namely, 
     no accuracy of arcseconds in Uhuru.

\item {\em Galactic coordinates}

\item {\em Reference for the position determination}: \\
     This is for readers who
      need to know the position accuracy and/or position determination method.
     Often SIMBAD gives the source position reference as well as the position itself, in which case we take both.   Otherwise, we searched for proper references by ourselves. 

\item {\em Spectral Model and Model Parameters}:  \\
We put a ``typical'' 
    spectral model and model parameters, including the
model normalization.  We follow the XSPEC (Arnaud 1996; http://xspec.gsfc.nasa.gov)
format for definition of the models and model parameters.
See Section \ref{SpectralModel} for details.

\item {\em Energy  fluxes}:\\
Energy fluxes are in ergs s$^{-1}$ cm$^{-2}$  and are
calculated from the assumed spectral models for several
energy bands.  We take typical JEMX 
soft and hard energy bands (3--10 keV and 10--30 keV),
and ISGRI soft and hard bands (20--60 keV and  60--200 keV). 

\item {\em JEMX, ISGRI  counting rates}:\\
Using the nominal detector responses, 
expected JEMX and ISGRI counting rates 
are calculated for the soft and hard energy bands indicated above.
In the actual QLA operation, 
JEMX and ISGRI source counting rates are 
measured in these energy bands to compare with the catalog.
The following calibration files, most recent ones in
July  2003 (but still based on pre-launch calibration),  
are used in this calculation: {\tt jmx2\_rmf\_grp\_0003.fits} for JEMX and
{\tt isgr\_rmf\_grp\_0005.fits} and
{\tt isgr\_arf\_rsp\_0002.fits} for ISGRI.

After the launch of INTEGRAL, the pre-launch estimate of the
JEMX effective area was revealed  to be 
overestimated by a factor of $\sim2$  (P.  Kretschmar, private communication).
Therefore, we divided the JEMX counting rates calculated using
the aforementioned responses by a factor of two.  This prescription 
has been confirmed to be valid 
from simultaneous INTEGRAL--XTE observations of 3C 273 (Courvoisier et al.\ 2003b) 
and Cyg X-1 (Pottschmidt et al.\ 2003).

For ISGRI, actual observations of Crab pulsar ane other sources
 confirmed that the pre-launch estimate
of the effective area was more or less correct (within $\sim 20$ \%
or so).

\item {\em Name aliases}:  \\
Since there are many varieties of sources,
      we do not have a consistent rule to decide which
      aliases to choose.  However, we try to put at least one 
      alias directly related to high-energy satellites and/or
      catalogs.  For example, ``4U'' for 4th Uhuru catalog
      (Forman et al.\ 1978), and ``2EG'' for second EGRET
      source catalog (Thompson et al.\ 1995).

\item {\em Comment on the nature of the source}:  \\
Such as low-mass X-ray 
binaries, high-mass X-ray binaries, Seyfert-1 galaxy, etc.

\end{itemize}

\section{Method to determine the  spectral models and fluxes}
\label{SpectralModel}

Except for  a small number of sources which are known to be invariable, 
it is hardly possible 
to put a single ``typical'' spectral model for each source 
to represent its  energy spectrum from 3 keV to 8000 keV,
In practice, however, we only require {\em rough}\/ estimates
of the source fluxes   for  the INTEGRAL QLA operation,
in which  we want to  detect  only the
very significant flux variations of more than a factor $\sim$10 with
JEMX and ISGRI in 3 to 200 keV.  This means that the requirement of the flux estimate in  our catalog
can be  as loose as a factor $\sim$10 in 3 to 200 keV, which we believe is
more or less achieved.

The published catalogs we use provide source fluxes in a very inhomogeneous set
of energy bands and units. In order to use this information in our catalog, it is
essential to harmonize the available data  in the published catalogs. 
For isolated pulsars (including the Crab pulsar) which are known to be invariable and well
approximated with power-law spectra, we adopt the known power-law slopes and 
normalizations. For other sources, 
we take a typical spectral
model   to describe the
emission for  a class of sources (Table 1), and  adjusted only  the normalization of the model 
for each source using the
published  data.
We use the spectral models that are appropriate for the most probable state of any given source
(for example, ``low'' state for Cyg X-1).  For X-ray novae, on the other hand, we use the
published fluxes obtained when the source is bright. 
Often there are two or more flux measurements for a given source in the same state;
in such a  case we adopt the average normalization which best fits all
the available measurements.

\begin{table*}
\begin{center}
\begin{tabular}{lcccccc}\hline\hline
Source category & model &$N_H$ &$\Gamma$1&$\Gamma$2&$E_{cut}$ &  $E_{fold}$ or  $E_{break}$$^{\dagger}$\\
                &       &($10^{22}$ cm$^{-2}$)  & & & [keV] & [keV]\\\hline
{\bf X-ray binaries}\\
LMXB&wabs*cutoff       &1.0   &1.7       & --    & 5 & --\\
LMXB with hard-tail&wabs*(cutoff+powerlaw)
                        &1.0   &1.7       &2.5    &10 & --\\
HMXB &wabs*highecut*powerlaw
                        &1.0   &1.0       &--     &10     &15\\
{\bf AGN}\\
Seyfert 1&wabs*cutoff  &1.0   &1.7       & --  &100 & --\\
Seyfert 2I&wabs*cutoff  &10.0  &1.7       & --  &100 & --\\
Blazar&wabs*cutoff      &1.0   &1.7       & --  &100 & --\\
Quasar&wabs*cutoff      &1.0   &1.7       & --  &100 & --\\
{\bf Clusters of Galaxies}&wabs*cutoff             &1.0   &1.7       & --  &10 & --\\
{\bf Unidentified}\\
X-ray sources&wabs*cutoff
                        &1.0  & 1.7       &--&10  & --\\
COMPTEL sources&wabs*bknpower
                        &100  &-1.5       &2.1  & -- & 1000\\
EGRET sources&wabs*bknpower
                        &100 &-1.5       &2.1  & -- & 10000\\\hline
\multicolumn{2}{l}{$^\dagger$$E_{fold}$ for highecut model and $E_{break}$ for bknpower model.}\\
\end{tabular}
\caption{Default spectral models adopted for different categories of sources.
See XSPEC manual (Arnaud 1996; http://xspec.gsfc.nasa.gov) for  precise descriptions
of the spectral models.}
\end{center}
\end{table*}

\section{Discussion}

In Figure 2, we show $\log N - \log S$ curves of the sources in our
catalog in 3 -- 10 keV and 20 -- 60 keV, as well as corresponding
JEMX and ISGRI counting rates and sensitivities.  The sensitivities
are calculated based on the pre-launch estimates available in the
IBIS and JEMX Observer's Manual, Issue 1.

Thanks to the large field of view and dithering strategy, INTEGRAL 
observations cover vast regions of the sky (IBIS bottom-to-bottom field of view is $29^\circ \times 29^\circ$).  
Thus the entire sky is expected to be covered during the INTEGRAL mission life,
at least with $\sim 10^5$ sec, which is a standard exposure time.
For the Galactic plane region, at least $10^6$ sec of exposure is expected because of the
Galactic center and plane survey core programs.  In Figure 2, we show 5 $\sigma$
detection limits for $10^5$ and $10^6$ sec exposures, respectively.
Since our catalog is not complete,  
we see flattening of the $\log N - \log S$ curves
at around the INTEGRAL detection limit,  more conspicuously
in JEMX.  In any case, our catalog includes $\sim 700$  
and  $\sim 400 $ sources, respectively, which are brighter than the 
JEMX and ISGRI sensitivities with $10^5$ sec exposure.
We may expect to detect still more sources, which may be 
transient, or dim persistent sources because of
the incompleteness of the catalog. Also, initial operation of ISGRI
has already detected several heavily absorbed, transient Galactic sources
(e.g, IGR J16318$-$4848 see Walter et al. 2003 and references therein),
which may not have been noticed previously below $\sim$10 keV, but
may be discovered with INTEGRAL.

Note that our catalog was made by combining observations of many different instruments;
 not a  single satellite has observed so many sources above 20 keV.
One of the main purposes of INTEGRAL is to create a homogeneous and 
most complete high-energy source catalog above 20 keV.

\begin{acknowledgements}
We acknowledge all the INTEGRAL team members who gave comments to 
improve early versions of the INTEGRAL reference catalog.
This research has made use of the SIMBAD database
operated at CDS (Strasbourg, France), NASA's Astrophysics Data System Service,
 the High Energy Astrophysics Science Archive Research Center (HEASARC) 
provided by NASA's Goddard Space Flight Center,
and the TARTARUS database supported by Jane Turner and Kirpal Nandra under NASA grants NAG5-7385 and NAG5-7067. 
We thank Dr. Peter Kretschmar for the information on the latest JEMX 
effective area calibration.
\end{acknowledgements}

   \begin{figure}
   \centering
   \includegraphics[angle=-90,width=8cm]{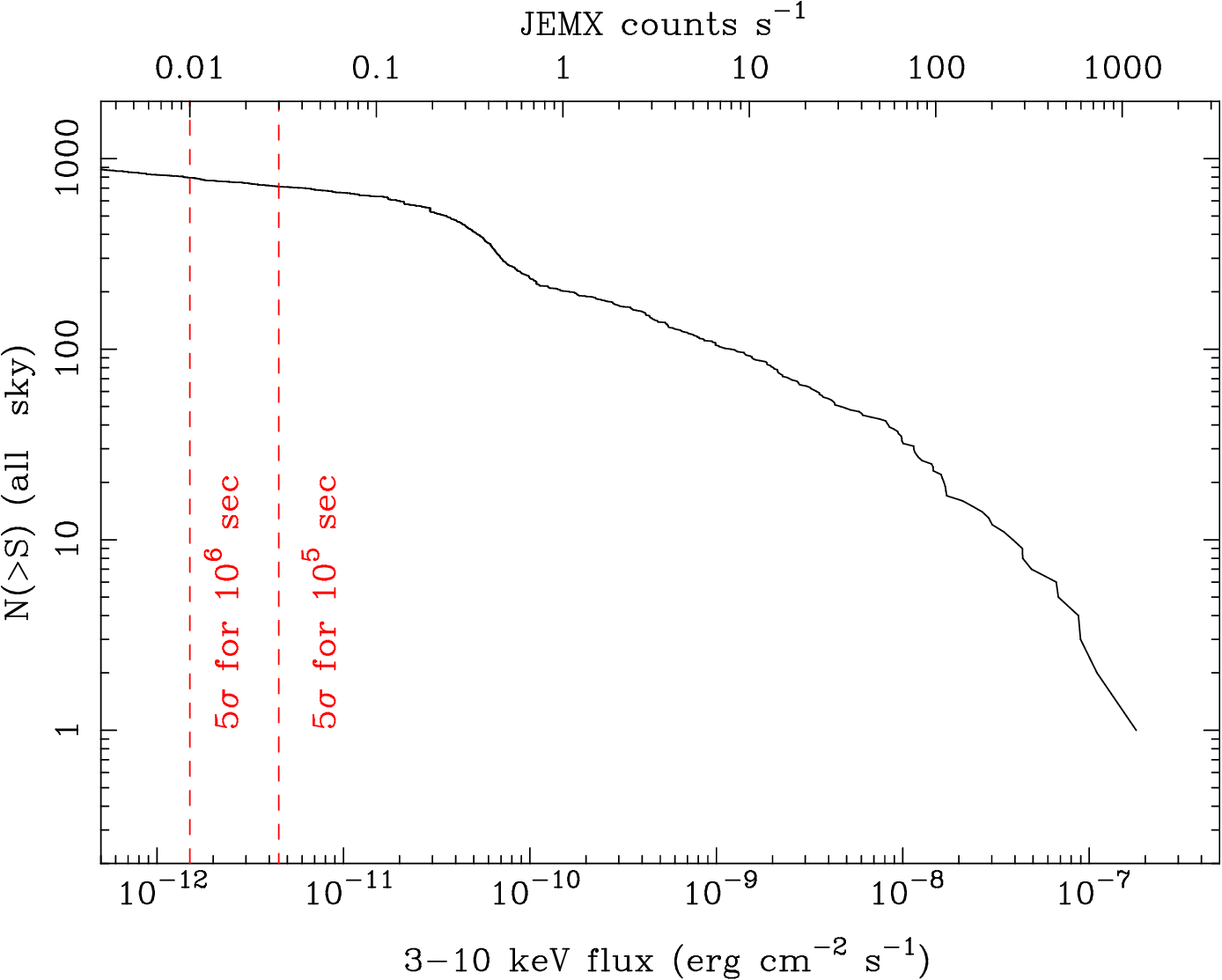}
   \includegraphics[angle=-90,width=8cm]{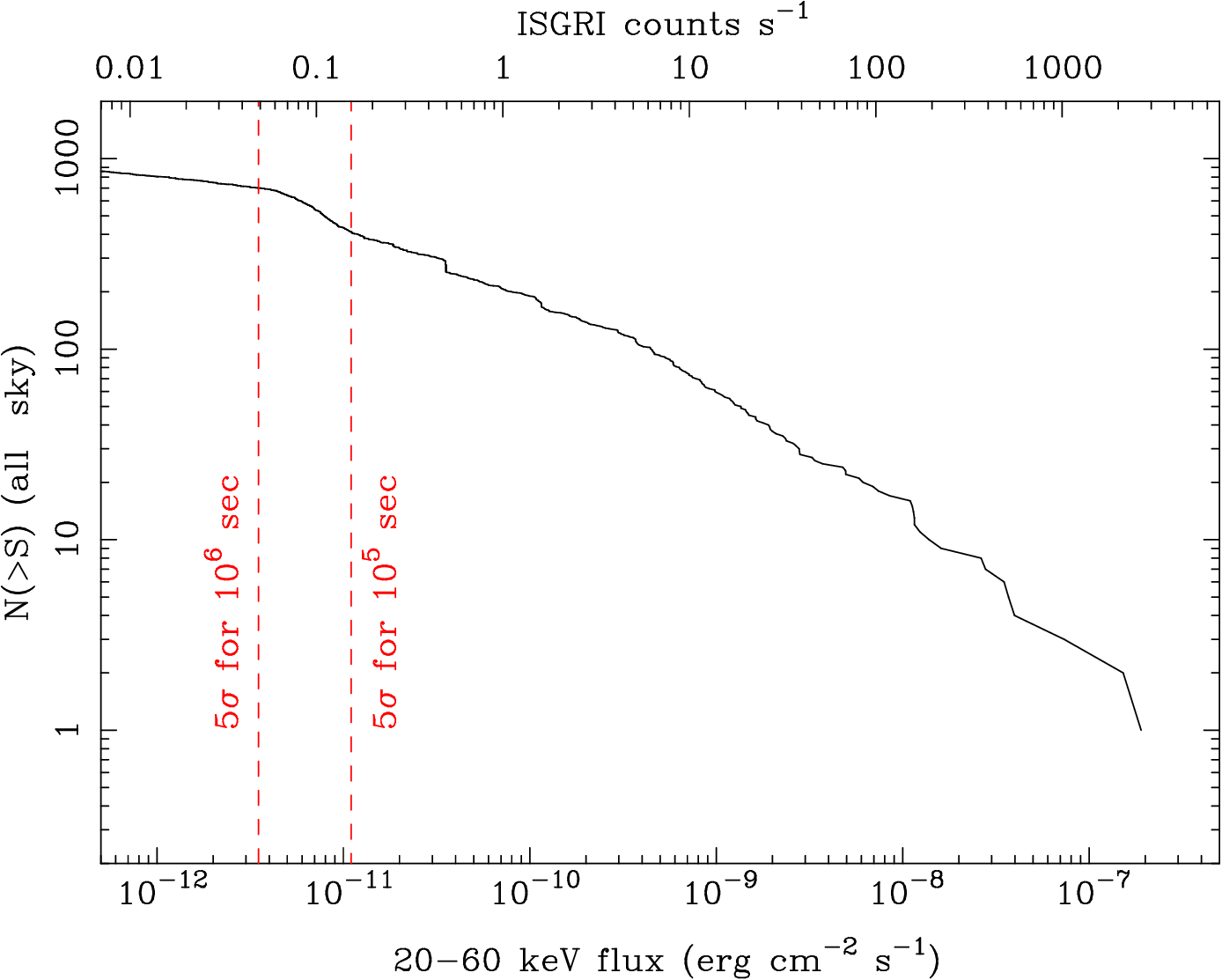}
      \caption{The $\log N - \log S$ curves of the sources in our catalog and
               JEMX and ISGRI sensitivities in the energy
               bands which are most sensitive and used for Quick Look Analysis
               (3 -- 10 keV for JEMX and 20 -- 60 keV
               for ISGRI).  JEMX and ISGRI counting rates for corresponding
               energy fluxes are also indicated.
              }
         \label{logNlogS}
   \end{figure}
\end{document}